\documentclass[prl,superscriptaddress,showpacs,showkeys,twocolumn]{revtex4-1}
\usepackage{amsfonts,amsmath,amssymb,graphicx,epstopdf,verbatim}
\usepackage[english]{babel}

\begin{document}

\title{The spatial coherence of weakly interacting one-dimensional non-equilibrium Bosonic quantum fluids}
\author{Vladimir N. Gladilin}
\affiliation{TQC, Universiteit Antwerpen, Universiteitsplein 1,
B-2610 Antwerpen, Belgium} \affiliation{INPAC,  KU Leuven,
Celestijnenlaan 200D, B-3001 Leuven, Belgium}
\author{Kai Ji}
\affiliation{TQC, Universiteit Antwerpen, Universiteitsplein 1,
B-2610 Antwerpen, Belgium}
\author{Michiel Wouters}
\affiliation{TQC, Universiteit Antwerpen, Universiteitsplein 1,
B-2610 Antwerpen, Belgium}
\date{\today}
\begin{abstract}
We present a theoretical analysis of spatial correlations in a
one-dimensional driven-dissipative non-equilibrium condensate.
Starting from a stochastic generalized Gross-Pitaevskii equation, we
derive a noisy Kuramoto-Sivashinsky equation for the phase dynamics.
For sufficiently strong interactions, the coherence decays
exponentially in close analogy to the equilibrium Bose gas.
When interactions are small on a scale set by the nonequilibrium
condition, we find through numerical simulations a crossover between
a Gaussian and exponential decay with peculiar scaling of the
coherence length on the fluid density and noise strength.
\end{abstract}

\maketitle

The spatial coherence is one of the key observables of quantum
degenerate Bose gases. While for Bose gases in thermal equilibrium, its
behavior is thoroughly understood \cite{lpss}, the case of
non-equilibrium Bose gases has received much less
attention. The non-equilibrium condition is typically of importance
for photonic systems, where due to the limited reflectivity of any
real mirrors, the photon life time is usually too short to achieve
true thermal equilibrium. A steady state arises instead thanks to the
balancing of external pumping and losses. Also in the
non-equilibrium situation, the spatial coherence remains a central
observable, that is experimentally accessible for photons
\cite{polbec}. Despite the existence of spatially extended lasers
for many decades \cite{lastrans}, the interest in the fundamental
properties of nonequilibrium quantum fluids has come quite recently,
with the advent of microcavity polariton quantum fluids
\cite{iac_review}. Polaritons are the quasi-particles that arise
from the strong coupling between a photon mode and an excitonic
excitation in a quantum well. They inherit the good coherence
properties from the photon together with substantial interactions
from the exciton. Experimentally, spatial coherence measurements
have been performed for two-dimensional polariton quantum fluids by
many groups \cite{polbec,coh_madr,coh_stan,coh_par,coh_sheff}.

In most experiments with microcavity polaritons, the nonequilibrium
condition is essential to understand their properties. It gives for
instance rise to a flow in the steady state
\cite{max_small,coh_par,prb2008}, that can assume the form of
quantized vortices \cite{vort} or feature more complicated patterns
\cite{baumberg}. From the theoretical point of view, these
nonequilibrium flows can be understood on a mean field level, from a
generalized Gross-Pitaevskii equation (gGPE)
\cite{prl2007,keeling2008}, that is of the same type as the complex
Ginzburg-Landau equation. In this Letter, we will consider an
equation in the form \cite{iac_review}
\begin{multline}
i\hbar \frac{\partial}{\partial t} \psi=
\left[-\frac{\hbar^2\nabla^2}{2m} +g |\psi|^2 + \frac{i}{2}
\left(\frac{P}{1+|\psi|^2/n_s}-\gamma \right) \right] \psi .
\label{eq:ggpe}
\end{multline}
Here $m$ is the effective mass and interactions are described by a
contact interaction with strength $g$. The imaginary term in the
square brackets on the right hand side describes the saturable
pumping (with strength $P$ and saturation density $n_s$), that
compensates for the losses ($\gamma$). The physical origin of the
pumping term for exciton-polariton condensates is an excitonic
reservoir that is excited by a nonresonant laser. For the case of
ordinary lasing in the weak coupling regime, it describes emission
of photons from the inverted electronic transition. In laser
physics, a wide variety of spatial optical patterns, described by
gGP-like equations, have been observed \cite{lastrans}.

In homogeneous systems, a simple uniform steady state solution of Eq.\eqref{eq:ggpe}
exists. When the pumping exceeds the losses ($P>\gamma$), it reads
$\psi_0=\sqrt{n_0}e^{-i\mu t/\hbar}$, where $n_0=n_s(P/\gamma-1)$
and the oscillation frequency is determined by the interaction
energy $\mu=g n_0$. For the description of the spatial coherence,
fluctuations have to be added to the gGPE. We will consider the
simplest case of spatially uncorrelated and spectrally white noise.
We thus supplement the right hand side of the gGPE with a stochastic
term $\sqrt{D} dW/dt$, where the complex stochastic increments
have the correlation function $\langle dW^*(x,t) dW(x',t') \rangle=2
\delta(x-x') \delta_{t,t'} dt  $. The coefficient $D$ describes
the strength of the fluctuations. The contributions from quantum fluctuations can for example be derived within a truncated Wigner approximation, leading
to $D\sim \gamma$ \cite{pol_twa}.

As usual for the calculation of the coherence of a low-dimensional
system, we perform the Madelung transformation $\psi=\sqrt{n}
e^{i\theta}$ to describe the complex field in terms of its density
and phase. This is most convenient, because the long-range part of
the spatial coherence is dominated by phase fluctuations. In the
regime of weak noise, the density fluctuations are small and we can
expand the interaction and gain saturation terms up to linear order
in $\delta n=n-n_0$. Because the low-momentum density fluctuations
relax much faster than the long wavelength phase fluctuations, they
can be adiabatically eliminated to derive an equation of the phase
dynamics only (see supplemental information for a derivation and a numerical check of the equivalence with Eq. \eqref{eq:ggpe}). It reads in Ito form
\begin{equation}
 d \tilde{\theta} =  \left[\tilde \mu  \nabla^2_{\tilde x} \tilde\theta
 -  \nabla^4_{\tilde x} \tilde \theta - ( \nabla_{\tilde x} \tilde \theta)^2 \right] d\tilde{t} + d \tilde W.
\label{eq:KSresc}
\end{equation}
Here, the increment $d\tilde W$ in Eq. \eqref{eq:KSresc}  is a real
stochastic variable with correlation function $\langle d \tilde
W(\tilde x,\tilde t) d\tilde W(\tilde x',\tilde t') \rangle =
\delta(\tilde x-\tilde x') \delta_{\tilde t,\tilde t'} dt $. We have
furthermore introduced the dimensionless variables $\tilde x=x/l_*$,
$\tilde t=t/t_*$ and rescaled the phase as $\tilde \theta =
\theta/\theta_*$. The length scale reads
\begin{equation}
l_*= \left(\frac{\hbar^2 }{2 m} \right)^{4/7} \eta^{3/7}
\left(\frac{D }{\hbar n_0} \right)^{-1/7} .
\end{equation}
Note its very weak dependence on the physical parameters $n_0$ and
$D$. The time and phase scales read,
\begin{eqnarray}
t_* &=& \hbar \left(\frac{\hbar^2 }{2 m} \right)^{2/7} \eta^{5/7}
\left(\frac{D }{\hbar n_0} \right)^{-4/7}, \\
\theta_* &=&
\left(\frac{\hbar^2 }{2 m} \right)^{-1/7} \eta^{1/7} \left(\frac{D
}{\hbar n_0} \right)^{2/7}.
\end{eqnarray}
They depend on the parameter $\eta$, that is a property of the gain
medium and can be written as $\eta=(1+n_s/n_0)\gamma^{-1}$, showing
that it decreases in our saturable gain model \eqref{eq:ggpe} for
increasing density $n_0$. The interaction energy, rescaled as $\tilde
\mu = \mu/\mu_*$ where
\begin{equation}
\mu_* = \frac{1}{2}\left(\frac{\hbar^2 }{2 m} \right)^{-1/7}
\eta^{-6/7} \left(\frac{D }{\hbar n_0} \right)^{2/7},
\end{equation}
remains as the only control parameter in the equation. Note that in
the absence of gain saturation $\eta \rightarrow \infty$ (obtained
in the limit $\gamma\rightarrow 0$), the time scale $t_*$ diverges.
The physical reason is that gain saturation is the only damping
mechanism in our model. We have verified that a small energy
relaxation term \cite{lp_damp} does not have a qualitative impact on
the results presented below.

The phase equation \eqref{eq:KSresc} is of the form of the noisy
Kuramoto-Sivashinsky equation (KSE) \cite{noisyKS}. The KSE equation
has a wide range applications, covering e.g.  flame fronts
\cite{siva} and reaction-diffusion systems \cite{kura}, as well as
transverse pattern formation in lasers \cite{kse_laser1,kse_laser2}.

The noisy KSE \eqref{eq:KSresc} has been studied in the context of
kinetic surface roughening \cite{cuerno_rough}. It has been shown
\cite{ueno} that it belongs to the universality class of the
celebrated Kardar-Parisi-Zhang (KPZ)   equation \cite{kpz} (Eq.
\eqref{eq:KSresc} without the fourth order derivative; for a review,
see \cite{halpin}). To the best of our knowledge, the spatial
correlations of the noisy KSE have however not been systematically
analyzed as a function of the control parameter $\tilde{\mu}$.

Let us start with the case of repulsive interactions $\tilde \mu>0$
(in equilibrium, this is the only meaningful case because attractive
interactions lead to a collapse) and defer attractive interactions
to the end. The linearized version of Eq. \eqref{eq:KSresc} is then
a good starting point to understand the momentum distribution. It
reads
\begin{equation}
\langle |\tilde \theta_{\tilde k}|^2\rangle = \frac{1}{2(\tilde{\mu}
\tilde k^2+\tilde k^4)}, \label{eq:momlin}
\end{equation}
where $\tilde k = k l_*$ is the rescaled momentum. The denominator
represents the square of the Bogoliuobov excitation spectrum $\tilde
\omega^2_B(k)= \tilde \mu \tilde k^2+\tilde k^4$.

The quadratic term dominates for wave vectors $\tilde k <
\sqrt{\tilde \mu}$. In this region, the phase dynamics reduces to
the KPZ equation, that was used for polaritons by
Altman {\em et al.} \cite{altmanKPZ}. Its relevance in the context
of equilibrium quantum fluids was also pointed out in recent works on
the dynamical structure factor of 1D bosons, both in the weakly
\cite{kulkarni} and the strongly \cite{gangardt_kpz} interacting
limits.

So far, the discussion is valid for arbitrary dimensionality, but
now we will restrict to a one-dimensional system, where the
nonlinear term in the KPZ equation leaves the solution of the
linearized equation invariant. The momentum distribution of the
phase field is then given by Eq. \eqref{eq:momlin}. This corresponds
to the quadratically decaying momentum distribution at long
wavelengths as in 1D finite temperature equilibrium condensates.
Eq. \eqref{eq:momlin} implies with the Fourier relation
\begin{equation}
\overline{\delta \theta^2}(x) =\frac{1}{L} \sum_{k} \, \langle |
\theta_{ k}|^2 \,\rangle  [\cos(kx)-1] \label{eq:four}
\end{equation}
that the phase-phase correlator decreases for long distances
linearly with the distance
\begin{equation}
\overline{\delta \tilde \theta^2} (x) \equiv \langle  \tilde
\theta(x) \tilde\theta(0) \rangle - \langle \tilde \theta^2(0)
\rangle  = - \tilde x/(4\tilde \mu),
\end{equation}

For the evaluation of the characteristic function that determines
the shape of the spatial coherence function $g^{(1)}(x)=\langle
\psi^\dag(x) \psi(0) \rangle $, the second cumulant approximation is
exact thanks to the Gaussian nature of the phase field in the
solution of the KPZ equation \cite{halpin}. For the correlation
length in the exponential decay $\exp\{i[\theta(x)-\theta(0)]\}
\rangle=\exp[ -\overline{\delta \theta^2}(x) ]$, we then recover the
expression for the coherence length $\ell_c=4 \hbar^2 n \eta \mu /(D
m)$, that was derived in Ref. \cite{al-iac}. An appealing
correspondence with the equilibrium case [$\ell_c^{\rm eq.}= 2 n
\hbar^2/(k_B T m)$] can be made by identifying the temperature $T$
with the noise strength and the mass with the `effective mass' $m
\eta/\mu$, that quantifies the curvature of the imaginary part of
the dispersion. A salient feature of this result is that the
coherence length tends to zero in the absence of interactions. The
linearized theory thus leads to the prediction that a repulsive
nonlinearity is essential for the spatial coherence of a
nonequilibrium condensate or a laser. The problem for vanishing
interactions can be immediately seen from Eqs. \eqref{eq:momlin} and
\eqref{eq:four}, because the sum over momenta features an infrared
divergence, due to the $k^{-4}$ divergence of the momentum
distribution.

In this case however, the nonlinear term in Eq. \eqref{eq:KSresc}
does not keep the momentum distribution \eqref{eq:momlin} obtained
in the linearized approximation invariant. We can expect on the
basis of a previous analysis \cite{ueno} of the noisy KSE that the
nonlinear term actually keeps the system in the KPZ universality class,
featuring a $k^{-2}$ behavior of the momentum distribution. In order
to investigate the spatial coherence for smaller interaction energy,
we have performed numerical simulations.

\begin{figure}[h!]
\centering
\includegraphics*[width=1\linewidth]{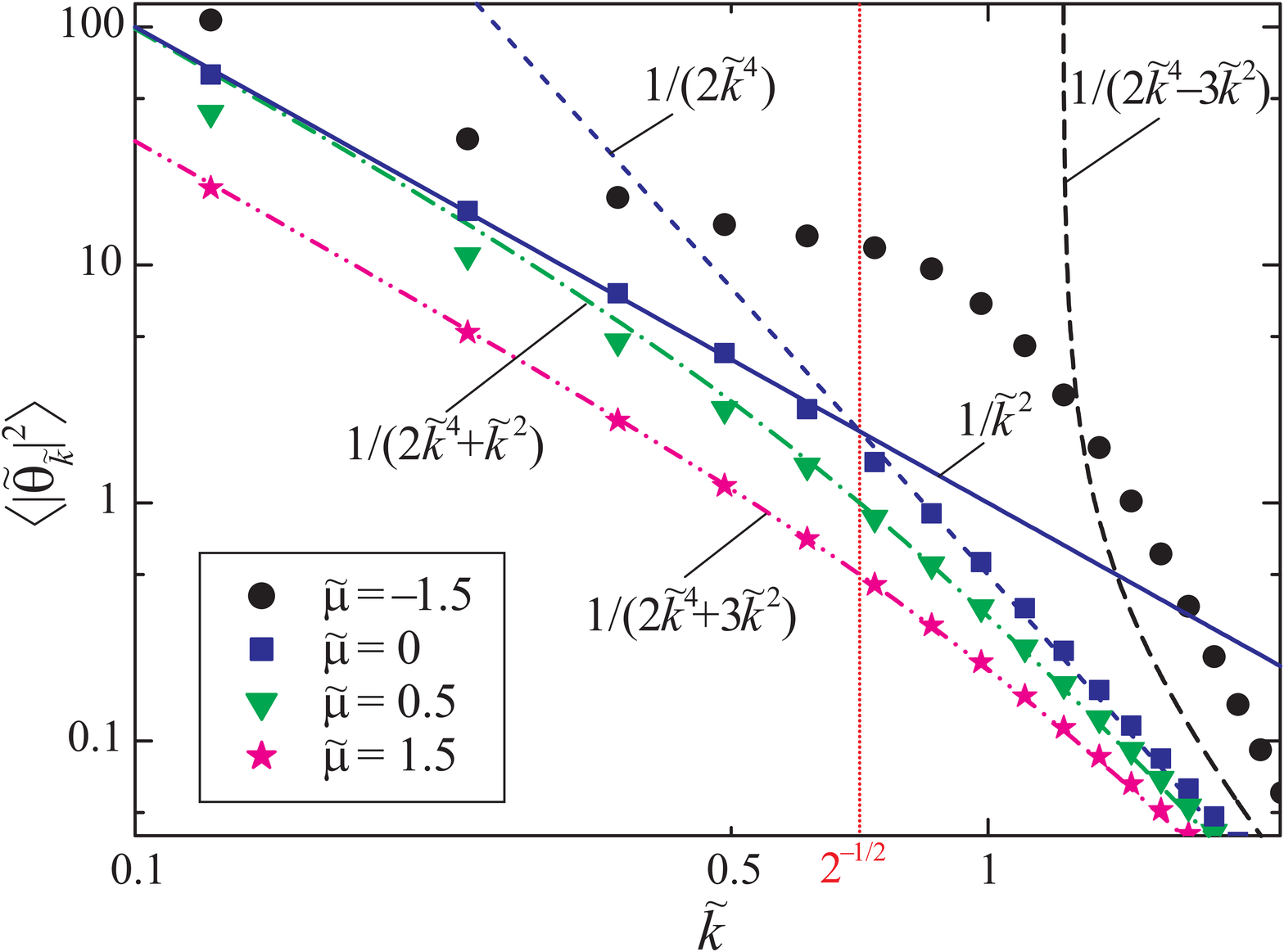}
\caption{Numerically calculated momentum distribution of the phase
field, $\langle |\tilde \theta_{\tilde k}|^2 \rangle$, as a function
of $\tilde k$ for $\tilde \mu=-1.5$ (circles), 0 (squares), 0.5
(triangles), and 1.5 (stars). The dashed, short-dash, dash-dot, and
dash-dot-dot curves represent the momentum distribution given by
Eq.~(\ref{eq:momlin}) for $\tilde \mu=-1.5$, 0, 0.5, and 1.5,
respectively. The function $\tilde k^{-2}$ (solid line) approximates
the calculated low-momentum behavior of $\langle |\tilde
\theta_{\tilde k}|^2\rangle$ in the absence of interactions. The
dotted vertical line corresponds to the crossover between this
behavior and a faster ($\propto \tilde k^{-4}$) decay at higher
momenta. \label{Fig1}}
\end{figure}

Figure~\ref{Fig1} present the momentum distribution of the phase
field for various values of the interaction energy. Our numerics
confirms the validity of the linearized approximation
\eqref{eq:momlin} for sufficiently strong repulsive interactions by
the excellent agreement between the stars and the dash-dot-dot line
for $\tilde \mu=1.5$. This corresponds to the KPZ regime discussed
above, where the nonlinearity of Eq. \eqref{eq:KSresc} does not
affect the static correlation function. For the intermediate
interaction strength on the other hand, pronounced differences
between the linearized approximation (dash-dotted) and the numerical
simulations (triangles) become apparent at low $k$.

For zero interaction strength, the discrepancy between the squares
and the dashed line is dramatic: instead of the $k^{-4}$ behavior
predicted by the linearized approximation, the momentum distribution
exhibits the same $k^{-2}$ divergence as in the KPZ interacting
regime. The prefactor in the low-momentum region turns out to be 1 within
our numerical accuracy of a few percent. As can be seen from Fig.
\ref{Fig1}, for $\tilde \mu=0$ the piecewise approximation to the
momentum distribution
\begin{equation}
\langle |\tilde \theta_{\tilde k}|^2 \rangle =
\left\{\begin{array}{c}
\tilde k^{-2} \;\;\;{\rm for} \; {\tilde k < 1/\sqrt{2}} \\
\\
{\frac{1}{2}\tilde k^{-4}} \;\;\;{\rm for} \; {\tilde k >
1/\sqrt{2}}
\end{array} \right.
\label{eq:mompiece}
\end{equation}
is excellent for almost all $\tilde k$, thanks to the sharp
crossover between the $\tilde k^{-4}$ and $\tilde k^{-2}$ regions.
This should be contrasted with the much smoother crossover --
described by the analytic function \eqref{eq:momlin} -- for strong
repulsive interactions (pink stars).

\begin{figure}[h!]
\centering
\includegraphics*[width=1\linewidth]{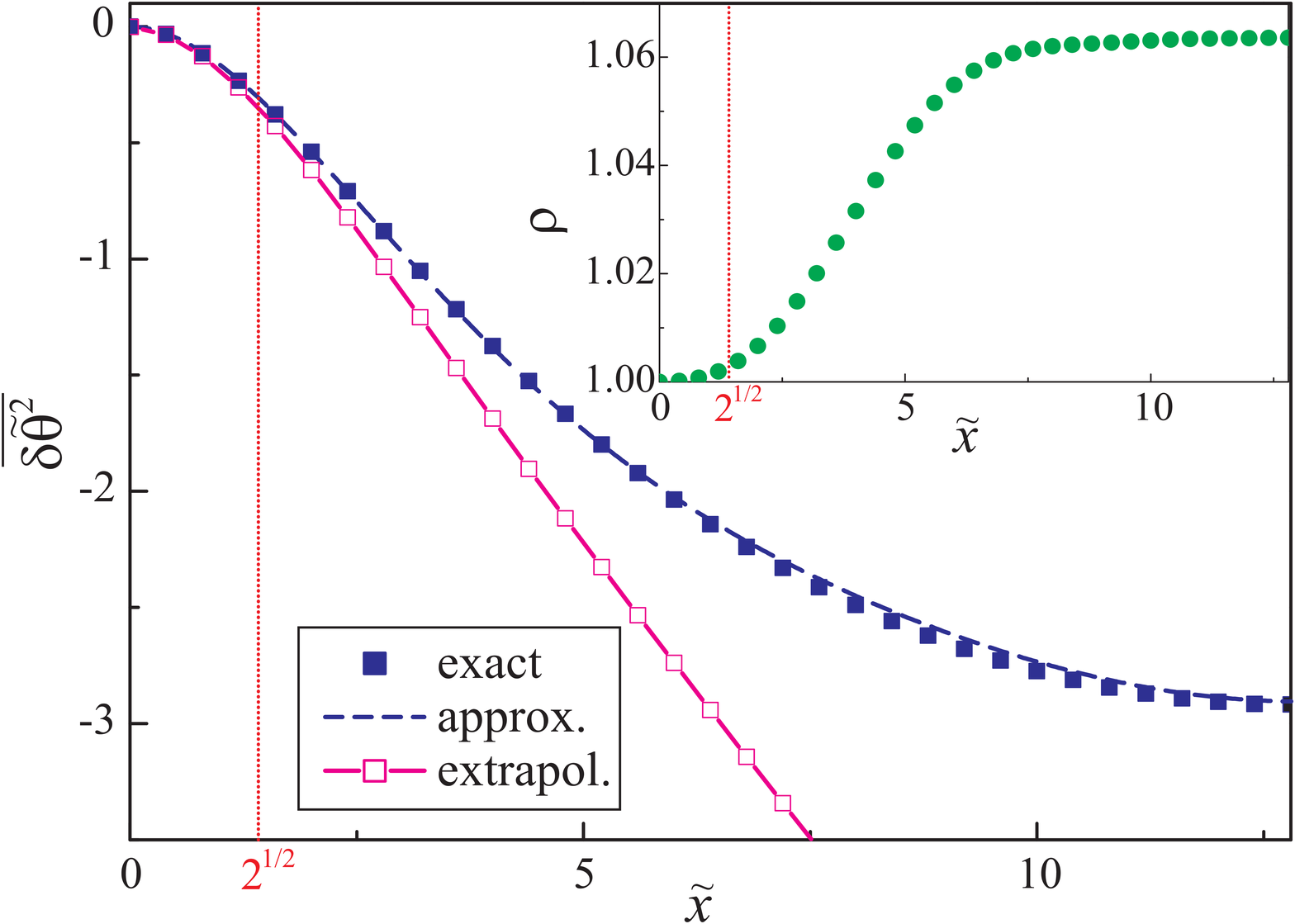}
\caption{Phase-phase correlator in real space for $\tilde \mu=0$,
$L=25.6l_*$ as given by the numerical simulations (full squares) and
by the analytical formula resulting from the piecewise approximation
\eqref{eq:mompiece}(dashed curve). The finite-size correction
described by this formula has been used to extrapolate the numerical
data to the case $L\to \infty$ (open squares). Inset: ratio $\rho(
x)=\ln \langle \exp\{i[\theta(x)-\theta(0)]\} \rangle
/\overline{\delta \theta^2}(x)$ calculated numerically for $\tilde
\mu=0$, $\theta_*=1$, $L=25.6l_*$. \label{Fig2}}
\end{figure}

The scaled phase-phase correlator in real space is shown in
Fig.~\ref{Fig2}.  The large distance ($\tilde x
\gg 1$) decay is dominated by the $\tilde k^{-2}$ region 
in the Fourier expansion \eqref{eq:four}  and reads
in the continuum limit $\overline{\delta \tilde \theta^2}(\tilde x)
= -\tilde x /2 $. The short range phase fluctuations are obtained by
expanding the cosine in Eq. \eqref{eq:four} up to second order. It
turns out that the low and high momentum regions contribute equally.
The piecewise momentum distribution \eqref{eq:mompiece} leads to
$\overline{\delta \tilde \theta^2}(\tilde x) =-\tilde x^2/\sqrt{2
\pi^2} $, valid for $\tilde x \ll 1$. Within the second cumulant
approximation, this leads to successive Gaussian and exponential
decays for the first order spatial coherence function $g^{(1)}(x)$.
The border between both regimes is determined by the length scale
$l_*$.

In the exponential regime, we find for the correlation length $l_e =
2 l_* \theta_*^{-2} $, featuring the peculiar scalings as a function
of the physical quantities $l_e \propto n_0^{5/7} D^{-5/7}
m^{-6/7}\eta^{1/7}$. Note the different scalings as a function of
the physical parameters as compared to the regime of strong
interactions. The Gaussian correlation length is $l_G=
(2\pi^2)^{1/4} l_*\theta_*^{-1}\eta^{2/7}$ and scales as $l_G
\propto n_0^{3/7} D^{-3/7} m^{-5/7}$. When the initial Gaussian
decay is by many orders of magnitude, the long distance exponential
tail becomes irrelevant. It is then meaningful to distinguish two
regimes. The decay is dominantly Gaussian when the `phase scale'
$\theta_*$ is much larger than one, where it is dominantly
exponential in the opposite limit. In terms of the physical
parameters of the nonequilibrium Bose fluid, the decay is dominantly
Gaussian for large noise and low density: $D > \hbar^2 n_0 / \sqrt{2
\eta m}$.

Thanks to the sharp crossover of the momentum distribution between
its two limiting behaviors, the piecewise approximation
\eqref{eq:mompiece} actually leads to an accurate expression for the
phase correlator at all distances. Its explicit form is given in the
supplemental material. In Fig.~\ref{Fig2}, it is displayed with a dotted line, showing almost perfect agreement with the numerical results. Furthermore,
the analytic formula allows us to do a straightforward extrapolation
to infinite systems (open squares in Fig.~\ref{Fig2}). The good
agreement between extrapolations starting from simulations with
different sizes (not shown here) confirms good convergence of our
results as a function of system size.

The above analysis was based on the assumption of the validity of the
second cumulant approximation for the characteristic function
$\langle \exp\{i[\theta(x)-\theta(0)]\} \rangle$. It is obviously
applicable for small phase variations, when this correlator is well
approximated by an expansion of the exponentials up to the second
order in $\theta$. For Gaussian phase fluctuations, it is also
guaranteed to hold, thanks to Wick's theorem. For the short
wavelength fluctuations, the nonlinear term can be neglected, so
that the phase increments are a linear combination of the input
noise, and therefore guaranteed to be Gaussian by the central limit
theorem. For the long wavelength components, that are crucially
affected by the nonlinearity, this argument breaks down. In order to
check the accuracy of the of the second cumulant approximation for the
characteristic function, we have evaluated it numerically. In the
inset to Fig.~\ref{Fig2}, we plot the ratio $\rho(x)=\ln \langle
\exp\{i[\theta(x)-\theta(0)]\} \rangle /\overline{\delta
\theta^2}(x)$ obtained for $\theta_*=1$. Starting from $\rho=1$ at
$ x\ll l_*$, this ratio slightly increases at $x\gtrsim \sqrt{2}l_*$
and apparently tends to saturate at even larger $ x$. The shown
behavior of $\rho(x)$ implies that the characteristic length for the
exponential decay of the correlator $\langle
\exp\{i[\theta(x)-\theta(0)]\} \rangle$ at long distances is about
six percent smaller than the value $l_e$, which follows from the
second cumulant approximation. It is worth mentioning that the
statistical convergence of the calculation of the correlator
$\langle \exp\{i[\theta(x)-\theta(0)]\} \rangle$ is much slower as
compared to the calculation of the phase fluctuations
$\overline{\delta \theta^2}$. Moreover, it is unfortunate that the
correction to the second cumulant approximation depends on the
parameter $\theta_*$ and hence on the strength of the noise, so that
one cannot obtain the correlator $\langle
\exp\{i[\theta(x)-\theta(0)]\} \rangle$ at different noise strengths
by a simple rescaling. On the other hand, as seen from the inset to
Fig.~\ref{Fig2}, even for the chosen value $\theta_*=1$, which
corresponds to a rather strong noise, the difference between the
exact correlator and the second cumulant approximation remains
moderate at all distances.

Finally, we briefly discuss the case of attractive interactions. The
equilibrium bose gas is unstable with respect to the collapse of the
gas because of the negative compressibility. The nonequilibrium
counterpart is reflected in the linear instability of Eq.
\eqref{eq:KSresc} for $\tilde \mu<0$. In the nonequilibrium case
however, the nonlinear term has a stabilizing effect, leading merely
to a modulation of the phase (and correspondingly of the density)
instead of a collapse. It is well known that for $\tilde \mu<0$, the
KSE without noise shows chaotic behavior, leading to effective
stochastic dynamics at large scales, where the stochastic driving
force stems from the linear instability \cite{halpin}. The
deterministic KSE with $\tilde \mu<0$ and the stochastic KPZ
equation belong to the same universality class \cite{lvov}. Adding a
stochastic term preserves this correspondence \cite{ueno}. The dots
in Fig. \ref{Fig1} show a momentum distribution for moderately
attractive interactions ($\tilde \mu =-1.5$). The bump around
$\tilde k=1$  also appears in the absence of the noise \cite{snep92}
and corresponds in real space to a modulation of the phase. At small
momenta, the momentum distribution shows the same $k^{-2}$ behavior
as in the case of repulsive or zero interactions, leading again to
an exponential decay at the longest distance scales.

In conclusion, we have shown that the coherence of a nonequilibrium Bose gas is governed by the noisy Kuramoto-Sivashinsky equation and we have analyzed the effect of interactions. In contrast to linearized Bogoliubov theory, our study has shown that the coherence length does not vanish when interactions tend to zero, even though it is shorter than in the case of repulsive interactions. Moreover, in our numerical calculations, we have observed a deviation from Gaussian statistics. Our predictions could be verified experimentally in semiconductor microcavities, where the repulsive interaction strength can be varied by changing the exciton-photon detuning \cite{polscatt}.

\newpage
\def\theequation{S.\arabic{equation}}
\setcounter{equation}{0}

\section{Supplemental material}

\subsection{Derivation of the phase equation}

Inserting $\psi=\sqrt{n_0+\delta n} e^{i\theta-i\mu t/\hbar}$ with
$|\delta n|\ll n_0$ into Eq.~(1), supplemented with a stochastic
term $\sqrt{D} dW/dt$, and neglecting terms of the order of $(\delta
n/n_0)^N$ with $N\geq 2$, one obtains
\begin{eqnarray}
\frac{\hbar}{2} \frac{\partial}{\partial t} \frac{\delta n}{n_0}&&=
-\frac{\hbar^2}{2m}\left[\nabla\frac{\delta n}{n_0} \nabla\theta
+\left(1+\frac{\delta
n}{n_0}\right)\nabla^2\theta\right]-\frac{1}{2\eta}\frac{\delta
n}{n_0}\nonumber\\
&&+ \sqrt{\frac{D}{n_0}}\frac{dW_n}{dt}, \label{dn}
\end{eqnarray}
\begin{eqnarray}
\hbar \frac{\partial}{\partial t} \theta &&=
\frac{\hbar^2}{2m}\left[\frac{1}{2}\nabla^2\frac{\delta n}{n_0}
-\left(\nabla \theta\right)^2\right]-\mu\frac{\delta n}{n_0}
\nonumber \\
&&+ \sqrt{\frac{D}{n_0}}\frac{dW_\theta}{dt}, \label{theta}
\end{eqnarray}
where the real stochastic variables $dW_n$ and $dW_\theta$ have
correlators $\langle dW_n(x,t) dW_n(x',t') \rangle= \langle
dW_\theta(x,t) dW_\theta(x',t') \rangle=\delta(x-x') \delta_{t,t'}
dt  $. Assuming that the characteristic time $\hbar \eta$, which
determines relaxation of density fluctuations, is sufficiently
short, the quantity $\delta n/n_0$ can be estimated from Eq.
(\ref{dn}) as
\begin{eqnarray}
\frac{\delta n}{n_0}\approx -
\frac{\hbar^2\eta}{m}\nabla^2\theta+2\eta
\sqrt{\frac{D}{n_0}}\frac{dW_n}{dt}.\label{dn2}
\end{eqnarray}
When inserting $\delta n/n_0$ given by Eq.~(\ref{dn2}) into
Eq.~(\ref{theta}), we take into account that at small $\eta$-values
under consideration (i) the inequality $\eta |\mu|\ll 1$ is
satisfied for a moderate interaction strength and (ii) the
contribution $\hbar^{2}\eta (2m)^{-1}\sqrt{D/n_0}\nabla^2(dW_n/dt)$
to the noise term can be neglected (except for the shortest length
scale, which we are not interested in). Then one obtains for the
phase equation the expression
\begin{eqnarray}
\hbar \frac{\partial}{\partial t} \theta &&=
\frac{\hbar^2}{2m}\left[-\frac{\hbar^2\eta}{2m}\nabla^4\theta
-\left(\nabla \theta\right)^2 +2\eta\mu \nabla^2\theta \right]
\nonumber \\
&&+ \sqrt{\frac{D}{n_0}}\frac{dW_\theta}{dt}, \label{theta2}
\end{eqnarray}
which immediately transforms into Eq.~(2) after substitutions
$x=l_*\tilde x$, $t=t_*\tilde t$, $\theta=\theta_*\tilde \theta $,
and $\mu_*\mu=\tilde \mu $ with $l_*$, $t_*$, $\theta_*$ and $\mu_*$
given by Eqs. (3) to (6). In the case of strong interactions, when
the inequality $\eta |\mu|\ll 1$ is violated, one should simply
replace the noise strength $D$ in Eq.~(\ref{theta2}) [as well as in
Eqs. (3) to (6)] with $D(1+2\eta \mu)^2$.

\subsection{Fitting formula for the phase-phase correlator}

We use the piecewise approximation of Eq.~(10) for the momentum
distribution $\langle |\tilde \theta_{\tilde k}|^2 \rangle$ in the
absence of interactions. Then for a finite system of (dimensionless)
size $\tilde L$  the scaled phase-phase correlator in real space can
be represented as
\begin{eqnarray}
\overline{\delta \tilde \theta^2}(\tilde x) &&\approx \frac{\tilde
L}{2\pi^2} \left[ \sum_{\kappa=1}^{\infty}\frac{\cos(2\pi \kappa
\tilde x/\tilde L)-1}{\kappa^2}\right. \nonumber
\\
&&\left.-\sum_{\kappa=\tilde
L/(2\sqrt{2}\pi)}^{\infty}\frac{\cos(2\pi \kappa \tilde x/\tilde
L)-1}{\kappa^2}\right]
 \nonumber
\\
&&+\frac{\tilde L^3}{(2\pi)^4}\sum_{\kappa=\tilde
L/(2\sqrt{2}\pi)}^{\infty}\frac{\cos(2\pi \kappa \tilde x/\tilde
L)-1}{\kappa^4}. \label{fit1}
\end{eqnarray}
Assuming that the system size is sufficiently large, $\tilde
L/(2\sqrt{2}\pi)\gg 1$, the last two sums in Eq.~(\ref{fit1}) can be
approximated by the corresponding integrals:
\begin{eqnarray}
\overline{\delta \tilde \theta^2}(\tilde x) &&\approx \frac{\tilde
L}{2\pi^2} \sum_{\kappa=1}^{\infty}\frac{\cos(2\pi \kappa \tilde
x/\tilde L)-1}{\kappa^2} \nonumber
\\
&&-\frac{\tilde x}{\pi}\int\limits_{\tilde x/\sqrt{2}}^{\infty}
dy\frac{\cos(y)-1}{y^2}
 \nonumber
\\
&&+\frac{\tilde x^3}{2\pi}\int\limits_{\tilde
x/\sqrt{2}}^{\infty}dy\frac{\cos(y)-1}{y^4}. \label{fit2}
\end{eqnarray}
These integrals as well as the infinite series in Eq.~(\ref{fit2})
can be easily evaluated~\cite{prud} giving finally
\begin{eqnarray}
\overline{\delta \tilde \theta^2}(\tilde x) &&=  -\frac{\tilde
x}{2}\left[1+\frac{2}{\pi} {\rm si}\left(\frac{\tilde
x}{\sqrt{2}}\right)\left(1+\frac{\tilde x^2}{12}\right)\right]
\nonumber \\
&&- \frac{2\sqrt{2}}{3\pi}\left[1-\cos\left(\frac{\tilde
x}{\sqrt{2}}\right)\left(1+\frac{\tilde x^2}{8}\right)\right.
\nonumber \\
&&\left.+\frac{\tilde x}{4\sqrt{2}}\sin \left(\frac{\tilde
x}{\sqrt{2}}\right)\right] +\frac{\tilde x^2}{2\tilde L},
\label{fit}
\end{eqnarray}
where $si(x)$ is the sine integral~\cite{abram}. The last term in
Eq.~(\ref{fit}) accounts for the finite size of the system.

\subsection{Impact of phase correlations on the field-field correlator}

\begin{figure}
\centering
\includegraphics*[width=1\linewidth]{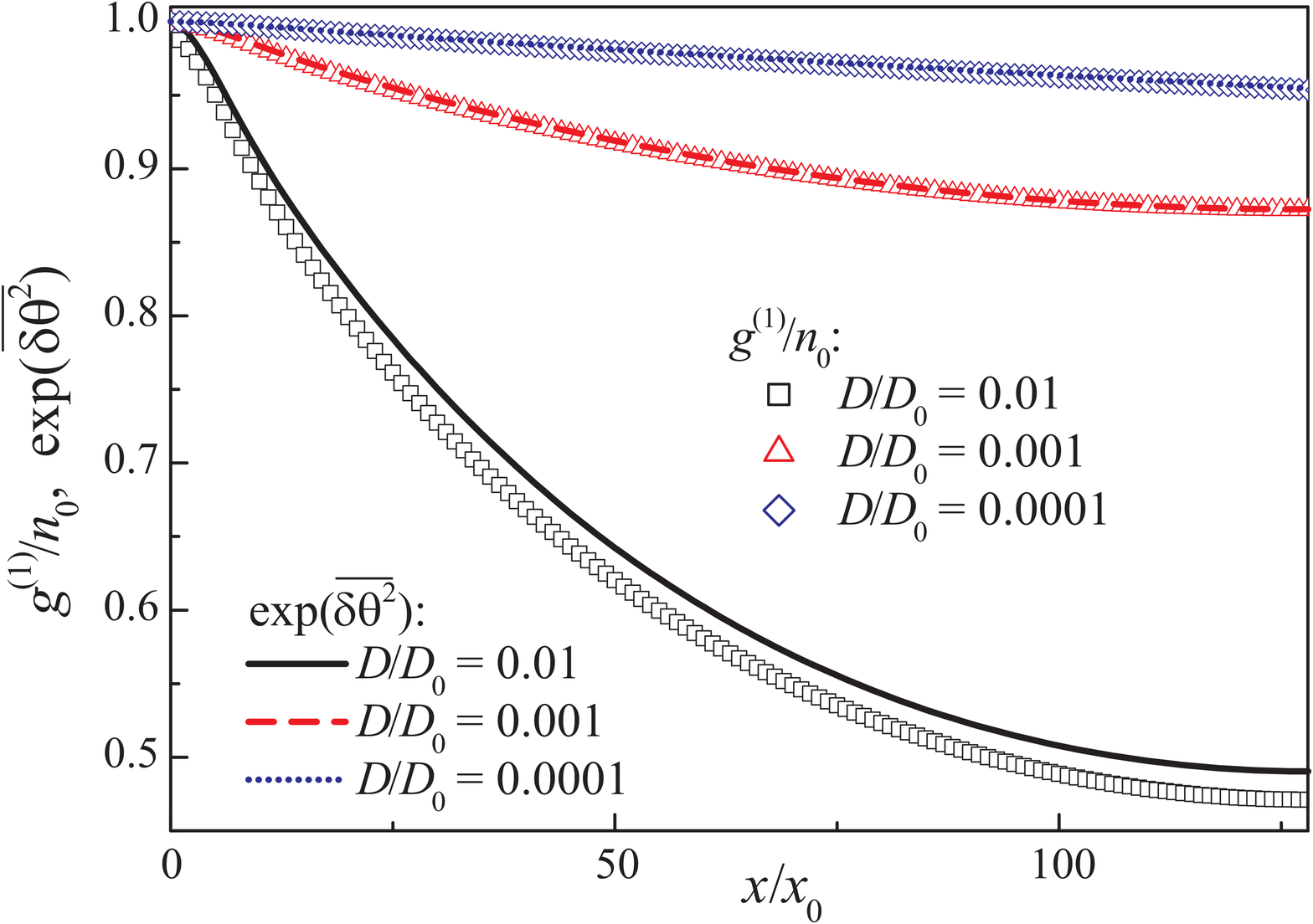}
\caption{First order spatial coherence function $g^{(1)}(x)$
calculated numerically from Eq. (1) (open symbols) and the function
$\exp[\overline{\delta \theta^2}(x)]$ (lines) for three different
values of the noise strength $D$. The system size is $L=1024 x_0$ in
the case of $D/D_0=0.0001$ and $L=256 x_0$ for the other two values
of $D/D_0$. Here $x_0=\sqrt{\hbar n_s/(m\gamma n_0)}$,
$D_0=\hbar^3n_0/(4mx_0)$. \label{FigS1}}
\end{figure}

In Fig.~\ref{FigS1} we compare the functions $\exp[\overline{\delta
\theta^2}(x)]$, obtained by rescaling the phase correlator
$\overline{\delta \tilde \theta^2}(\tilde x)$ (see the paper), to
the the first order spatial coherence function $g^{(1)}(x)/n_0$,
calculated directly from Eq.~(1), for the case of $g=0$, $n_0=n_s$
and three different values of the the noise strength $D$. As follows
from Fig.~\ref{FigS1}, while at weak noise (up to $D/D_0 \sim
10^{-3}$) the behavior of the field-field correlator $g^{(1)}(x)$ is
perfectly described by the phase factor $\exp[\overline{\delta
\theta^2}(x)]$, at higher noise strength an additional suppression
of $g^{(1)}(x)/n_0$ by density fluctuations becomes non-negligible.
Nevertheless, even in the case of $D/D_0 = 10^{-2}$ where the
magnitude of density fluctuations is comparable to $n_0$, the decay
of $g^{(1)}(x)/n_0$ at large distances is seen to be dominated by
the effect of phase fluctuations.

\begin{figure}

\centering
\includegraphics*[width=1\linewidth]{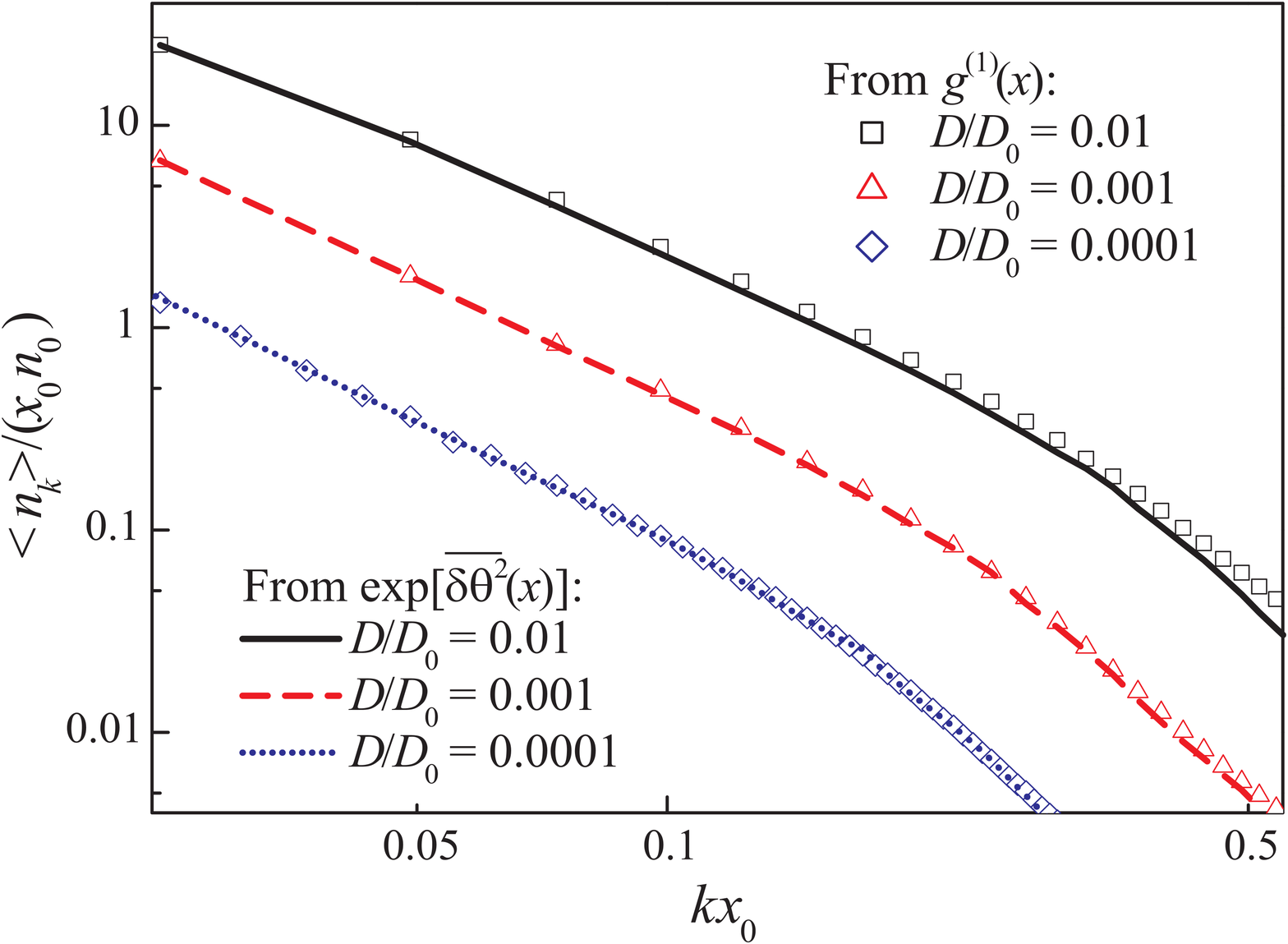}
\caption{Momentum distribution functions calculated from the
filed-field correlator $g^{(1)}(x)$ (open symbols) and from the
functions $\exp[\overline{\delta \theta^2}(x)]$ (lines) for three
different values of the noise strength $D$. The system size is
$L=1024 x_0$ in the case of $D/D_0=0.0001$ and $L=256 x_0$ for the
other two values of $D/D_0$. Here $x_0=\sqrt{\hbar n_s/(m\gamma
n_0)}$, $D_0=\hbar^3n_0/(4mx_0)$. \label{FigS2}}
\end{figure}

In Fig.~\ref{FigS2} we plot the momentum distribution functions,
which correspond to $g^{(1)}(x)$ (open symbols) and
$\exp[\overline{\delta \theta^2}(x)]$ (lines). In line with the
discussion above, the long-wavelength behavior of the momentum
distribution obtained by solving the ``full'' field equation (1) is
completely determined by phase fluctuations. The effect of
(short-range) density fluctuations is manifested only for relatively
large momenta at sufficiently high noise strength.

\end{document}